\documentclass[aps,
               pra,
               twocolumn,
               nofootinbib,
               floatfix,
               showpacs
              ]{revtex4}

\usepackage{amsmath, amssymb}
\usepackage{graphics}
\usepackage{graphicx}
\usepackage{color}

\renewcommand{\i}{\mathrm{i}}

\renewcommand{\vector}[1]{\mathbf{#1}}

\newcommand{\eq}[1]{(\ref{eq:#1})}
\newcommand{\Eq}[1]{Eq.~(\ref{eq:#1})}

\newcommand{\Fig}[1]{Fig.~\ref{fig:#1}}

\begin{document}
\title{Tuning universality far from equilibrium}

\author{Markus Karl}
\author{Boris Nowak}
\author{Thomas Gasenzer}
\email{t.gasenzer@uni-heidelberg.de}
\affiliation{Institut f\"ur Theoretische Physik,
             Ruprecht-Karls-Universit\"at Heidelberg,
             Philosophenweg~16,
             69120~Heidelberg, Germany}
\affiliation{ExtreMe Matter Institute EMMI,
             GSI Helmholtzzentrum f\"ur Schwerionenforschung GmbH, 
             Planckstra\ss e~1, 
            64291~Darmstadt, Germany} 

\date{\today}

\begin{abstract}
Possible universal dynamics of a many-body system \emph{far} from thermal equilibrium are explored. 
A focus is set on meta-stable non-thermal states exhibiting critical properties such as self-similarity and independence of the details of how the respective state has been reached. It is proposed that  universal dynamics far from equilibrium can be tuned to exhibit a dynamical phase transition where these critical properties change qualitatively. This is demonstrated for the case of a superfluid two-component Bose gas exhibiting different types of long-lived but non-thermal critical order. Scaling exponents controlled by the ratio of experimentally tuneable coupling parameters offer themselves as natural smoking guns. The results shed light on the wealth of universal phenomena expected to exist in the far-from-equilibrium realm.
\end{abstract}


\maketitle

%
%
The concept of universality has been extremely successful in classifying and characterising equilibrium states of matter. 
For example, there are different types of order in a magnetic material separated by a second-order phase transition at which the relevant physical properties become independent of the microscopic details of the system. 
This constitutes universality and allows to characterise an extensive range of different phenomena in terms of just a few classes governed by the same critical properties. 
In view of the intensifying discussion on the dynamics of many-body systems it is a pressing question whether also far away from the thermal limit the character of dynamical evolution can become independent of the microscopic details. 
For a closed system this would imply that in approaching a critical configuration the evolution must become independent of the particular initial state the system has started from and critical slowing down in the actual time evolution is observed. 
As a consequence, different types of dynamical evolution could be distinguished by means of  universality classes.

\begin{figure}[!t]
    \begin{centering}
    \includegraphics[width=0.5\textwidth]{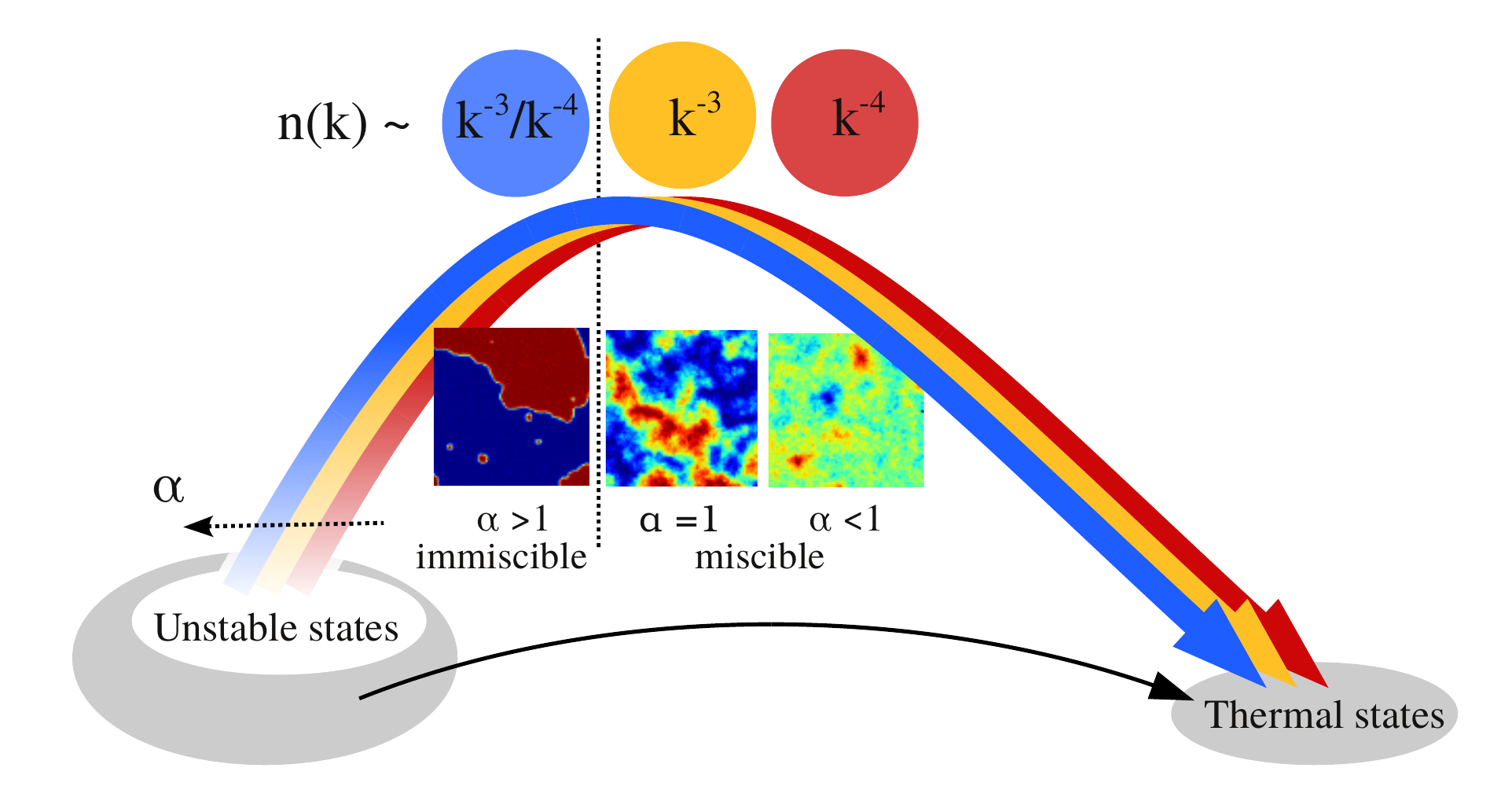}
    \end{centering}
    \caption{Illustration of the dynamical evolution, with a focus on the possible transition between far-from-equilibrium transient states. 
    The set in the bottom left corner represents an ensemble of out-of-equilibrium initial states.
    Amongst these, a subset subjected to a dynamical instability is marked in white.
    It is chosen such that for all values of the the ratio $\alpha$ of inter- to intra-species couplings the system belongs to the unstable subset. 
    Subsequently, the time evolution of the closed system leads to non- and quasi-topological defect formation. 
    The type of defects created determines the scaling in the low-momentum regime of the particle spectrum of the gas. 
    Our results reveal three different scalings corresponding to three types of defect configurations, and a dynamical phase transition between the far-from-equilibrium universal states. 
    }
    \label{fig:One}
    \end{figure}

In this article, we demonstrate that such universal dynamics far from thermal equilibrium states is indeed possible. 
We show that there is a dynamical phase transition between different types of universal dynamics. 
After having quenched a two-component ultracold Bose gas we follow the ensuing evolution of the closed system and observe transient universal behaviour. 
In particular, we identify a tunable external parameter that determines the type of transient non-equilibrium order. 
In our setup non-equilibrium order is constituted by the appearance of spatial patterns including quasi-topological defects like vortices and skyrmions which are created through instabilities and which decay only on very large time scales~\cite{Nelson2002a}. 
We find the dynamics of diluting defects to constitute a separate dynamical critical phenomenon far from thermal equilibrium and therefore clearly distinct from the known equilibrium critical points~\cite{Hohenberg1977a}. 
Large-scale correlations in the system are universal in the sense that they are fixed by the type of defect, but are completely insensitive to the specific positions and velocities of the defects. 
Distinguishable types of defects are produced for different values of the tuning parameter. 
Hence, the dynamics can be tuned to a transition between different metastable non-equilibrium ordered states, 
see \Fig{One}.  

%
%
%
    \begin{figure*}[!t]
    \includegraphics[width=0.95\textwidth]{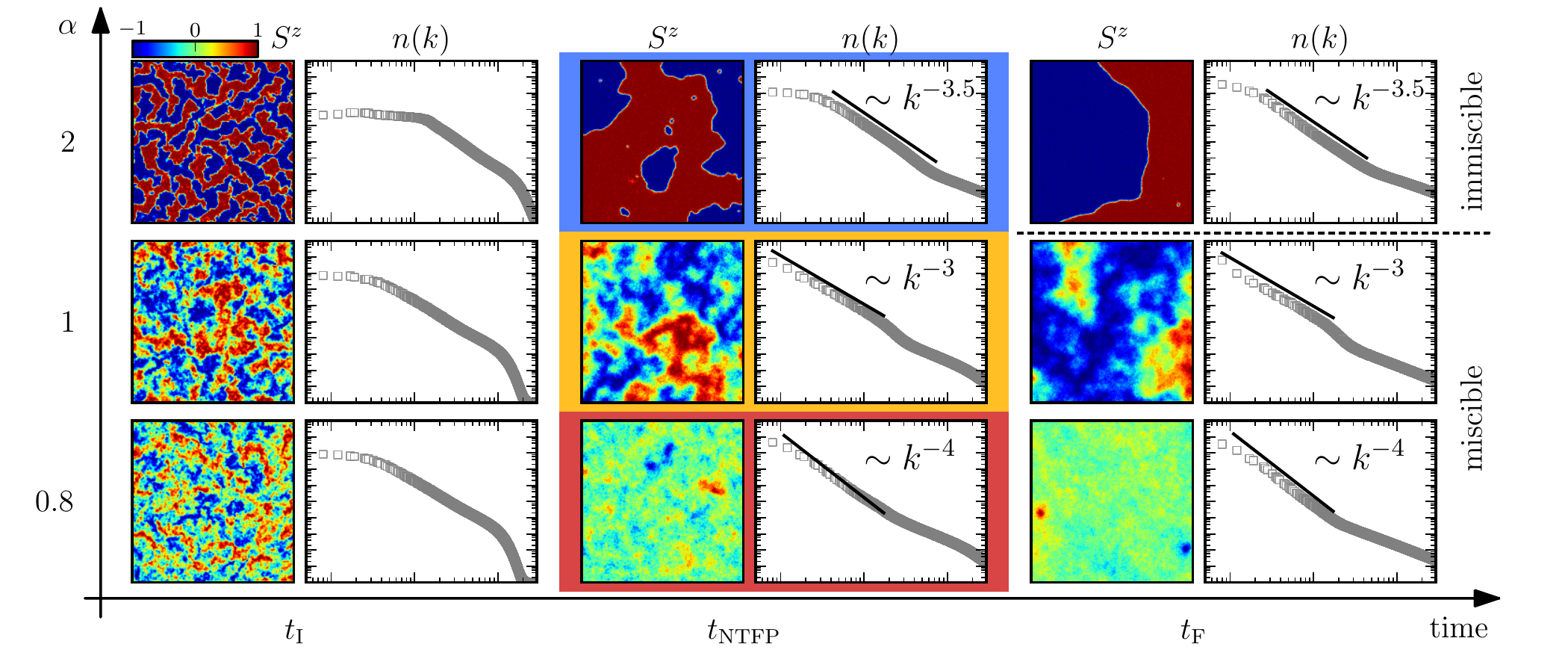}
    \caption{Three snapshots (columns) of the evolving two-component Bose gas for three different values of the coupling ratio $\alpha$ (rows). 
In each case we show the spatial distribution of density imbalance $S^z$ as well as the momentum distribution $n(k)$ as a function of the momentum modulus $k$ of particles in the gas. 
Tick labels for the spectra graphs are the same as in \Fig{Three}. 
Until $t_\mathrm{I}$ the dynamics is characterised by the instability and isotropisation of density fluctuations. 
At $t_\mathrm{NTFP}=10\,t_\mathrm{I}$, one observes the development of different power laws in the momentum distributions. 
They remain metastable for a long time beyond $t_\mathrm{F}=100\,t_\mathrm{I}$ and reveal different non-thermal fixed points.
The change of power laws at $\alpha=1$ indicates a dynamical phase transition.}
    \label{fig:Two}
    \end{figure*}

Extending upon the presented results will allow to learn about the universal properties of classes of models in the vicinity of attractors, non-thermal fixed points and other critical submanifolds of the greater space of all possible non-equilibrium configurations~\cite{Berges:2008wm, Bonini1999a, Berges:2008sr, Scheppach:2009wu}. 
Knowing such universal properties, predictions for the behavior of very different physical systems can be obtained on the basis of comparatively few exemplary measurements. 
When looking at fundamental science applications, universality classes of far-from-equilibrium criticality can link the dynamics, which we propose to be observable in ultracold atomic or exciton-polariton gases, with phenomena as different as, \textit{e.g.}, magnon gases in solids, quark-gluon-plasma dynamics in heavy-ion collisions, or reheating after early-universe inflation.
Technological applications would not stay away.
 
The two-component Bose systems we consider are described by the Hamiltonian density ($\hbar = 1$) 
%
\begin{align}
\label{eq:hamiltonian}
  \mathcal{H} 
  =  \frac{1}{2m}\nabla\phi_j^{\dagger}\nabla\phi_j + \frac{g}{2}{(\phi_j^{\dagger}\phi_j)}^2 - g(1-\alpha)\phi_1^{\dagger}\phi_1\phi_2^{\dagger}\phi_2\,.
\end{align}
%
Here $m$ is the mass of the atoms and $\alpha = g_{12}/g$ the ratio between the inter-species coupling $g_{12}$ and the intra-species interaction constant $g$. 
The latter as well as the mass are 
chosen to be the same for both components and the sum over the field index $j \in \{1,2\}$ is implied.  
The considered systems allow for good experimental control and have been studied intensely~\cite{Hall1998a, Guzman2011a, Sabbatini2011a, Nicklas2011a}. 
In experiment, $\alpha$ can be varied by means of a Feshbach resonance. 
Two different ground states exist, depending on the value of the parameter $\alpha$~\cite{Timmermans1998a, Kasamatsu2006a}. 
In the immiscible regime, $\alpha > 1$, the inter-species interaction energy is greater than the intra-species interactions. 
Hence, in the ground state, the spatial overlap of the components is minimised through domain formation and spatial separation of the two components.
On the contrary, for $\alpha \leq 1$, the two components become miscible and, in the ground state, uniformly distributed over the whole volume.
We make use of $\alpha$ to change the properties of the system in the yet unexplored region of non-equilibrium quasi-stationary states. 
A focus is set on long-lived states with non-/quasi-topological defects including domain walls~\cite{Timmermans1998a, Hall1998a}, single-species vortices, and skyrmions~\cite{Kasamatsu2005a} in the coupled system.

    \begin{figure*}[!t]
    \includegraphics[width=0.95\textwidth]{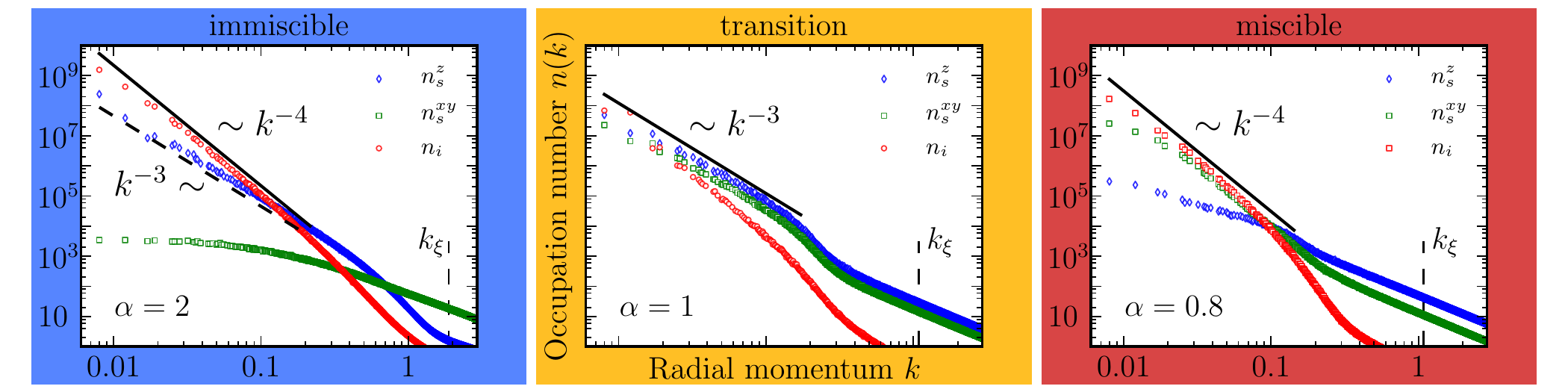}
    \caption{Decomposition of correlation functions of the two-component Bose gas in the three different non-equilibrium ordered states parametrised by the coupling ratio $\alpha$. 
Shown are the $z$-spin component $n_s^z$, the sum of $x$- and $y$-spin components $n_s^{xy}$ and the incompressible component $n_i$ at time $t_\mathrm{NTFP}$.
For details of the decomposition procedure see~\cite{Supplement}.
}
    \label{fig:Three}
    \end{figure*}

Since the dynamics we are interested in exclusively affects the low-momentum, strongly populated field modes, we employ the so-called classical field method which yields, within numerical accuracy, exact results for the time-evolving observables. 
In order to discuss the contribution of the different (quasi-)topological configurations in detail, we make use of the spin representation of the two-species fluid. 
In this representation the $S^z$-component of the spin vector simply refers to the density imbalance between the two species. 
This is an essential observable for the detection of domain walls. 
Furthermore, we make use of a hydrodynamic decomposition of the superfluid flow which allows for the detection of vortex contributions to the spectra. 
For details on numerical parameters and decomposition methods we refer the reader to the supplementary material~\cite{Supplement}.

In \Fig{Two}, we unravel the time evolution of a two-component Bose gas as outlined in \Fig{One}. 
The parameters of our simulation are chosen such that the final states are close to the ground state of the system. 
To clarify the type of non-equilibrium order during the intermediate stages of the evolution, we show the imbalance between the two components $S^z$ as well as the momentum distribution $n(k)$ for three characteristic times. 
The initial time, $t_\mathrm{I}$, marks the stage of an isotropic momentum distribution which is overpopulated within an intermediate range of momenta, as compared to the ensuing equilibrium distribution. 
It is marked by a strong fall-off at large momenta. In our driving scheme this state is reached in the wake of an instability. 
For the immiscible case, $\alpha>1$, we use a modulational instability in which small low-momentum fluctuations in the polarisation are amplified to macroscopic spin domains. 
In the miscible regime, $\alpha\leq 1$, we invoke a counter-superflow instability ~\cite{Takeuchi2010a, Takeuchi2011a} by choosing oppositely directed flow vectors for the two field components. 
Above a critical relative velocity, momentum exchange between the two fluids causes the superflow to decay and spin domains to form.
Subsequently, overpopulation at intermediate momenta is encountered for all values of the parameter $\alpha$. 
Its microscopic origin is seen in the coloured distributions of the density imbalance $S^z$ in \Fig{Two} which show strong fluctuations and the onset of domain formation. 
We estimate $t_\mathrm{I} \simeq 10/\omega_\mathrm{I}$, with $\omega_\mathrm{I}$ being the energy of the fastest growing unstable mode. 
Note, that the dynamics up to $t_\mathrm{I}$ is different for different $\alpha$ and involves for example isotropisation in case of $\alpha\leq 0$~\cite{Supplement}. 
The further dynamical evolution of overpopulated states involves particle transport towards small momenta and energy transport towards high momenta~\cite{Nowak:2010tm, Nowak:2011sk}. 
In \Fig{Two} this is best observed by comparing the high momentum tails of the momentum distributions at $t_\mathrm{I}$ and $t_\mathrm{NTFP}$. 
Particle transport towards low momenta eventually fills the zero-mode (not seen), a process most important in the miscible regime $\alpha=0.8$. 
The key insight gained from the spectra concerns the development of infrared (IR) power laws in the momentum distribution $n(k) \sim k^{-\zeta}$ at $t_\mathrm{NTFP}\simeq 10^2/\omega_\mathrm{I}$. 
These power laws  depend on the external parameter $\alpha$: $\zeta\sim3.5$ for $\alpha>1$, $\zeta\sim3$ for $\alpha=1$ and  $\zeta\sim4$ for $\alpha<1$. 
Also, the pattern of imbalance fluctuations in the system depends on the interaction strength. 

We highlight the most striking observations at late times before we carry on developing a detailed understanding of the relation between the dominant fluctuations and the observed power laws. 
In the immiscible regime the system consists of few large domains with additional  inclusions of small point-like domains. 
As compared to the initial time domains have considerably grown. 
A special situation is encountered at the transition point ($\alpha=1$), where domain-like structures persist to extremely long times. 
This is remarkable since in this regime domains are  not energetically favourable as compared to overlapping particle densities. 
For $\alpha=0.8$, imbalance fluctuations have decayed up to few small areas of strong imbalance. 
Let us finally look at the largest computed time  $t_\mathrm{F}\simeq 10^3/\omega_\mathrm{I}$. 
For $\alpha>1$, two domains of equal size remain, which reflects the immiscibility in the ground state. 
The small-scale domain-fluctuations have considerably reduced in number. 
At the transition point, we observe the persistence of domain-like structures which we attribute to a diverging time scale $\tau$ of domain decay as $\alpha \rightarrow 1$ from below~\cite{Supplement}. 
For $\alpha<1$,  small long-lived imbalance fluctuations remain, whereas the background tends to become very smooth, $S^z \approx 0$. 

Finally, we investigate the microscopic origin of the scaling found in the momentum distributions. 
The bimodal power laws in these distributions are signatures of the system having approached a non-thermal fixed point~\cite{Berges:2008wm, Berges:2008sr, Scheppach:2009wu}. 
At long times they become more pronounced in all three cases, signalling critical slowing down of the dynamic evolution.
In \Fig{Three}, the result of a decomposition of the momentum distribution according to spin and fluid degrees of freedom at $t=t_\mathrm{NTFP}$ is presented~\cite{Supplement}. 
We show spin-spin correlation functions as well as correlations in the incompressible velocity as a function of momentum. 
The crucial point is that they explain the scaling properties of the total momentum distribution and allow for an interpretation in terms of specific defect configurations, as shown in \Fig{Two}. 
This completes our understanding of transient non-equilibrium order in a two-component ultracold Bose gas.

We remark that, for all $\alpha$, compressible and quantum pressure excitations, which are not shown, dominate the spectrum in the ultraviolet (UV) regime and follow a thermal Rayleigh-Jeans distribution $n \sim k^{-2}$, while they give a negligible contribution towards lower momenta. 

In the immiscible regime, see \Fig{Three} (left), the main contribution to the spectrum in the IR region is provided by the incompressible component $n_i$, corresponding to flow orientations transverse to the direction in which the flow velocity varies. 
Although we are dealing with a multi-component gas this feature is similar to superfluid turbulent flow in a single-component Bose gas. 
Thereby the incompressible spectrum shows $n_i \sim k^{-4}$ scaling over approximately one decade which is generated by coherent vortical flows $\vector{w}_i$ around topological defects~\cite{Supplement, Nowak:2010tm, Nowak:2011sk}. 
The cores of these vortex-like structures can be seen in the spin imbalance, see \Fig{Two}, since they are filled with particles of the other species and thus are of the skyrmion type. 
They are created during the merging process of  domains and, persisting due to their topological nature, give the main contribution to the incompressible component spectrum. 
However, spin excitations $n_s^z$ overtake in an intermediate momentum region, showing a scaling of $n_s^z \sim k^{-3}$. 
Looking more closely, one observes that the scaling behaviour terminates in the IR at a scale $\pi/L_\mathrm{D}$ given by the mean domain size $L_\mathrm{D}$, while the cut-off in the UV at $\pi/\xi_s$ is set by the width of the domain walls, \textit{i.e.}, the spin healing length $\xi_s =\xi (1 / |1-\alpha|)^{{1}/{2}}$. 
Since the two contributions $n_s^z$ and $n_i$ are of comparable magnitude within an intermediate momentum range the sum of all contributions, giving the full spectrum $n(k)$, appears to follow the power law $n \sim k^{-3.5}$ in the IR, as discussed above. 

The situation on the other side of the transition can also be clarified. When $\alpha$ falls below $1$ domains are energetically suppressed. Thus, vortices dominate the non-thermal fixed point and induce the characteristic scaling $n \sim n_i \sim k^{-4}$, see \Fig{Three} (right). Here, it is also visible in the scaling of relative phase fluctuations $n_s^{xy}$, which is related to constant particle densities on scales larger than $\xi$ as compared to the skyrmionic case above. 

The picture changes dramatically on the transition point, $\alpha=1$. 
As shown in \Fig{Three} (middle), vortical flow is much less important in this regime. 
The momentum distribution is dominated by spin fluctuations scaling as $n \sim n_s^z \sim n_s^{xy} \sim k^{-3}$~\cite{Supplement}. 
We attribute this feature to the increased dimensionality of the vacuum manifold for $\alpha=1$ which removes the topological protection of vortices. 
The stability of domain walls at the transition point takes over but is non-topological in nature. 
Instead, conservation laws restrict the decay of this particular type of defect~\cite{Lee1992a}. 
This argument can be made even more transparent by studying the critical dynamics at the transition from $\alpha<1$ towards $\alpha=1$, where two non-equilibrium ordered states meet each other, vortices and domains~\cite{Supplement}. 

We conclude that it is possible to establish the notion of a dynamical phase transition between different far-from-equilibrium ordered states in a two-component Bose gas. 
Beyond that, the underlying concept of \mbox{(non-)}topological defects determining bulk features of correlation functions in far-from-equilibrium situations is very general. 
It is easily imaginable that multi-component field theories with more than two components show similar behavior to the one outlined here. 
We point out the possibility of a universal duality between decaying defects and inverse particle cascades. 
This requires the generation of (quasi-)topological configurations far from thermal equilibrium and their slow decay, going together with an increase of coherence and defect separation~\cite{Schole:2012kt}. 
Under these conditions, an inverse particle cascade is generated, and the associated power-laws can be found from the scaling properties of the respective single defect. 
A variety of (quasi-)topological excitations are known to exist in multi-component fields~\cite{Nelson2002a, Lee1992a}, examples are monopoles in gauge fields~\cite{Rajantie2002a} and exotic magnets~\cite{Castelnovo2008a}, as well as skyrmions in Bose-Einstein condensates~\cite{Ruostekoski2001a,Kasamatsu2005a} and liquid crystals~\cite{Dierking2003a}. 
New interesting features that are readily accessible in experiment are expected for ultracold spinor gases~\cite{Stenger1999a, Sadler2006a, Vengalattore2008a, Ueda2012a, Fujimoto2012a}. 
The transition between different types of transient non-equilibrium order can be controlled by changing the symmetry properties of the Hamiltonian and thus topology and local conservation laws of the system. 
This offers interesting prospects for far-from-equilibrium dynamical phase transitions in very different areas of physics.

\textit{Acknowledgements}.
We thank J. Berges, S. Diehl, S. Erne, P. Kevrekidis, L. McLerran, E. Nicklas, M. Oberthaler, J. M. Pawlowski, J. Schole, D. Sexty, and C. Wetterich for useful discussions. 
Work supported by Deutsche Forschungsgemeinschaft (GA677/7,8), University of Heidelberg (CQD), and Helmholtz Association (HA216/EMMI).

\newpage
\mbox{}
\newpage


\section{Supplementary Material}
\label{sec:SuppMat}
In the supplementary material we provide details of the methodology as well as additional results supporting our approach and conclusions.
%
\subsection{Classical field equations}
\label{subsec:Model}

Since the dynamics we are interested in exclusively affects the low-momentum, strongly populated field modes, we employ the so-called classical field method which yields, within numerical accuracy, exact results for the time-evolving observables~\cite{Blakie2008a, Polkovnikov2010a}. 
For this, initial field configurations $\phi_{1,2}(x,t_0)$ are sampled from Gaussian probability distributions and then propagated according to the classical equations of motion. 
At the end of the time evolution correlation functions are obtained from ensemble averages over the set of sampled paths. 
The classical equations of motion derived from the Hamiltonian density \eq{hamiltonian} of the interacting two-component Bose gas are

%
      \begin{subequations}
      \label{eq:GP}
	\begin{align}
	\mathrm{i}\partial_t\phi_1 &= -\frac{1}{2}\nabla^2\phi_1 + g(|\phi_1|^2 + \alpha|\phi_2|^2)\phi_1 \label{subeq:GP-1}\,,\\
	\mathrm{i}\partial_t\phi_2 &= -\frac{1}{2}\nabla^2\phi_2 + g(|\phi_2|^2 + \alpha|\phi_1|^2)\phi_2 \label{subeq:GP-2}\,.
	\end{align}
      \end{subequations}
%

The momentum distribution which is shown in various graphs is defined as   
%
	\begin{equation}
	\label{eq:sinpartspec}
	n(k,t) = \int \! \mathrm{d}\Omega_k \, \langle \phi_j^{\ast}(\vector{k},t)\phi_j(\vector{k},t) \rangle\,,
	\end{equation}
%
where $\int \! \mathrm{d}\Omega_k$ denotes the angular average in two-dimensional momentum space. 
                               
We have applied the following rescalings to obtain dimensionless quantities in the system of coupled Gross-Pitaevskii equations \eq{GP}, using the lattice constant $a_s$ of the computational grid: $\phi a_s \to \phi$, $mg \to g$ and $t/(m a_s^2) \to t$. 
Here $n = (N_1 + N_2)/L^2$ is the mean total particle density on a $N_s\times N_{s}$ grid of linear size $L=N_s a_s$, with $N_s=1024$. 
Conversion to physical time scales appropriate for a specific experiment, with, $\emph{e.g.}$, $^{87}\mathrm{Rb}$ atoms, can be achieved by noting that the healing length $\xi = (2mgn)^{-{1}/{2}}$ is given by $\xi=4a_s$ for our system. 
Giving an estimate, \emph{e.g.}, for a physical healing length $\xi \simeq 1\,\mu$m, our lattice unit corresponds to $a_s \simeq 0.25\,\mu$m which yields a time unit of $m_\mathrm{Rb}a_s^2/\hbar \simeq 10^{-4}\,$s. 
Hence, the characteristic time scale is $1/\omega_\mathrm{I} \simeq 10^{-3}\,...\,10^{-2}\,$s which  yields a realistic time scale for the appearance of the non-thermal fixed point at $t_\mathrm{NTFP}\simeq 0.1\,...\,1\,$s.

 \subsection{Overpopulation from dynamical instability}
\label{subsec:Overpop}
%
%
    \begin{figure}[!t]
    \includegraphics[width=0.45\textwidth]{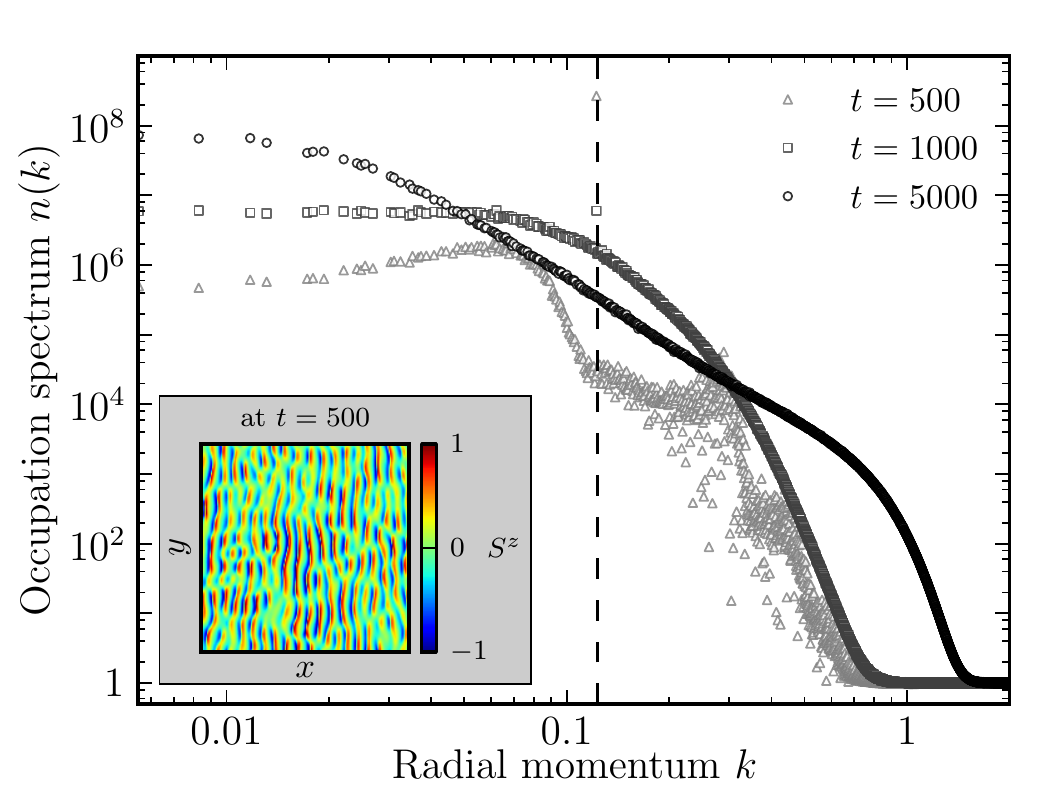}
    \caption{Momentum distribution $n(k)$, defined in \Eq{sinpartspec}, at different grid times for the case $\alpha=1$. 
The counterflow mode is indicated by the dashed vertical line. 
The inset figure shows a snapshot of a typical spin configuration $S^z$, corresponding to the earliest depicted spectrum at grid time $t = 500$. 
$t = 5000$ corresponds to $t_\mathrm{I}$ discussed in \Fig{Two} in the main text.}
    \label{fig:supp_fig1}
    \end{figure}
%
In our scheme the exponential growth of low lying momentum modes is the first step towards the approach of the non-thermal fixed point. 
The production of an overpopulated state via instabilities has been discussed in \Fig{Two} of the main text. 
Here, we review the initial dynamics in more detail for the case of $\alpha=1$, see~\Fig{supp_fig1}. 
The counter-superflow instability leads to a growth of certain unstable modes, acting as a source for low momenta and depleting the counterflow mode. 
This initiates isotropisation and a typical cascading process in momentum space, causing the transport of particles to the lower momentum modes while their energy is carried away by a few particles moving to the higher momentum modes. 
\Fig{supp_fig1} shows this evolution until the earliest time shown in \Fig{Two} in the main text when universal features start to dominate the dynamic evolution. 

%
%
\subsection{Spin-fluid representation}
\label{subsec:SpinRep}
Although we always simulate the full set of equations \eq{GP} it is instructive to study the degrees of freedom which describe the relative evolution of the two components. 
Writing the fields in the polar representation $\phi_i = \sqrt{\rho_i}\exp{(\i\theta_i)}$ these are given by the local phase difference $\theta_r = \theta_1-\theta_2$ and by the local density difference $\rho_1 - \rho_2$. 
The Pauli matrices $\sigma^{a}$ serve to define the Schwinger representation $S^a = \phi_i^{\dagger}\sigma_{ij}^a\phi_j$ (sum over repeated indices implied) of angular momentum operators.
This results in a three-component vector of (pseudo-)spin densities ${S^a}$, $a \in \{x,y,z\}$
%
      \begin{subequations}
	\label{eq:SpinRep}
	\begin{align}
	 S^x &= 2\sqrt{\rho_1\rho_2}\cos{(\theta_r)} \label{subeq:SpinRep-x}\,,\\
	 S^y &= -2\sqrt{\rho_1\rho_2}\sin{(\theta_r)} \label{subeq:SpinRep-y}\,,\\
	 S^z &= \rho_1 - \rho_2 \label{supeq:SpinRep-z}\,,
	\end{align}
      \end{subequations}
%
where the modulus corresponds to the total density $|\vector{S}| = \rho_1 + \rho_2 \equiv \rho_T$. 
For convenience, we apply the redefinition $S^a \to \rho_TS^a$ such that $|\vector{S}| \equiv 1$. 
Then, the Hamiltonian density $\mathcal{H}$ of \Eq{hamiltonian} can be rewritten as 
%
	\begin{align}
	\label{eq:energy}
	\nonumber  \mathcal{H} 
		&= \frac{1}{2}(\nabla\sqrt{\rho_T})^2 + \frac{\rho_T}{8}\nabla S^a  \nabla S^a + \frac{1}{2\rho_T}\vector{j}_T^2 \\
		&~~+ \frac{g\rho_T^2}{2} -\frac{g\rho_T^2}{4}(1-\alpha)\left[1-(S^z)^2\right]\,,
	\end{align}
%
which shows the influence of the relative degrees of freedom and their coupling to the global ones~\cite{2Kasamatsu2005a}. 
The quantity $\vector{j}_T = \rho_1\nabla\varphi_1 + \rho_2\nabla\varphi_2$ is the conserved total particle current associated with a global $U(1)$ symmetry of \Eq{hamiltonian} associated with the global shift of the total phase $\theta_T = \theta_1 + \theta_2$. 
Thus $\vector{j}_T$ can not be expressed using just the spin densities but contains also the remaining fourth degree of freedom, the total phase $\theta_T$. 
We obtain
%
	\begin{equation}
	\label{eq:totcurr}
	\vector{j}_T = \frac{1}{2}\rho_T\nabla\theta_T + \frac{\rho_TS^z}{2\left[1-(S^{z})^{2}\right]}(S^y\nabla S^x - S^x\nabla S^y)\,.
	\end{equation}
%
According to the representation \eq{energy} of the Hamiltonian density the two-component Bose gas can be described by a spinor field $\vector{S}$ which is coupled to a hydrodynamic fluid with density $\rho_T$ and (conserved) quasi particle current $\vector{j}_T$.
For a fluid at rest, \textit{i.e.}, $\rho_T = \mathrm{const}$ and $\vector{j}_T = 0$, the spin system thereby assumes the form of a classical non-linear sigma model (NL$\sigma$M) with a mass term $g\rho_T^2/4(1-\alpha)\left[(S^x)^2+(S^y)^2\right]$,
whereas in general a current $\vector{j}_T \neq 0$ leads to a highly non-trivial coupling between internal and hydrodynamic degrees of freedom. 

%
%
 \subsection{Hydrodynamic decomposition}
\label{subsec:Hydrodecomp}
In order to gain further insight into the interplay between the domain structure and other excitations we decompose the energy spectrum according to Refs.~\cite{Nore1997a, Nore1997b,2Nowak:2011sk}. 
The results of this decomposition at the non-thermal fixed point were discussed in \Fig{Three} of the main text. 
In a density-phase representation, the kinetic part of the Hamiltonian density can be written in the following way:
%
	\begin{align}
	\label{eq:energydec}
	 \mathcal{H}_{\mathrm{kin}} &=  \nabla \phi_j \nabla \phi_j 
			  = |\nabla\sqrt{\rho_T}|^2 + \frac{\rho_T}{4}\nabla S^a  \nabla S^a + |\vector{w}|^2 \,.
	\end{align}
%
The velocity field $\vector{w}$ is defined via the total particle current $\sqrt{\rho_T}\vector{w} = \vector{j}_T$, similar to the convenient choice for the one-component case. 
In this decomposition the first and the last terms are the quantum-pressure component and the classical hydrodynamic component, respectively. 
In contrast to a single-component fluid the second term of \Eq{energydec} adds a pressure-like contribution to the kinetic energy which is produced by internal excitations only. 
In addition, the velocity field $\vector{w}$ and thus the corresponding part of the kinetic energy can be decomposed in a compressible and an incompressible part, $\vector{w} = \vector{w}_c + \vector{w}_i$, with $\nabla \times \vector{w}_c = 0$ and $\nabla \cdot \vector{w}_i = 0$, such that wave-like and vortical excitations appear in different parts of the decomposition, see Refs.~\cite{Nore1997a, Nore1997b}. 
Based on \Eq{energydec}, radial energy spectra in momentum space corresponding to the discussed energy parts can be defined as
%
	\begin{subequations}
	\label{eq:radenergyspec}
	  \begin{align}
	      E_{\delta}(k) &= \frac{1}{2} \int \! \mathrm{d}\Omega_k \, \langle |\vector{w}_{\delta}(k)|^2 \rangle,~~~\delta \in \{q,c,i\} \label{subeq:rad-1} \, \\
	      E_{s}(k) &= \frac{1}{2} \int \! \mathrm{d}\Omega_k \, \langle \vector{w}_{s}^a(k) \cdot \vector{w}_{s}^a(k)  \rangle \,. 
\label{subeq:rad-2}  
	  \end{align}
	\end{subequations}
%
Here we have introduced additional velocities $\vector{w}_q = \nabla\sqrt{\rho_T}$ and $\vector{w}_s^a = \sqrt{\rho_T}\nabla S^a/2$ for the sake of a closed representation. 
Finally, the
energy spectra can be converted to occupation number spectra defined in \Eq{sinpartspec} by multiplication with a factor $k^{-2}$, $n_{\delta}(k) = k^{-2}E_{\delta}(k)$, $\delta \in \{q,c,i,s\}$~\cite{2Nowak:2011sk}.

%
    \begin{figure}[!t]
    \includegraphics[width=0.45\textwidth]{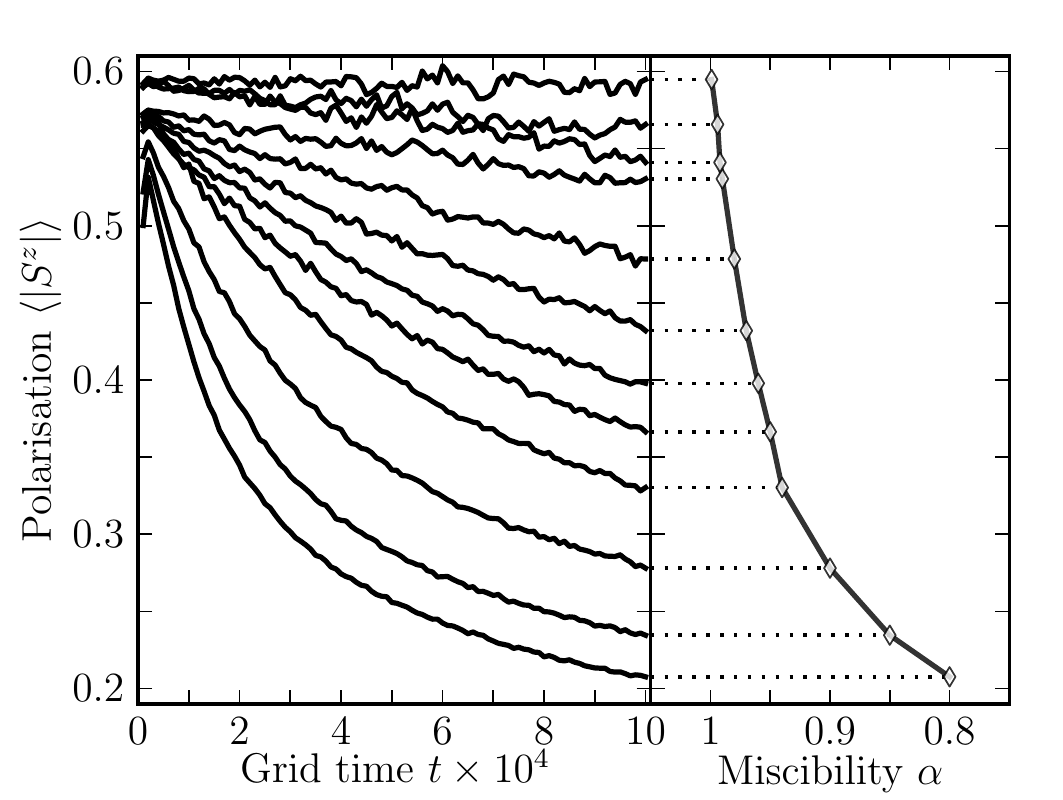}\\
    \includegraphics[width=0.45\textwidth]{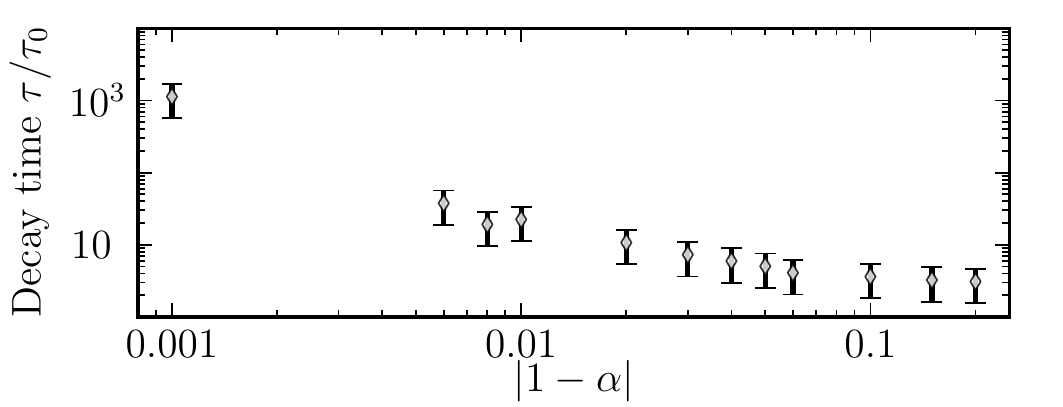}
    \caption{Top left: Integrated density imbalance or polarisation $\langle |S^z| \rangle$ as a function of time for various $\alpha$. 
Top right: Final values of $\langle |S^z| \rangle$ as a function of $\alpha$, showing a slow down of the imbalance decay when $\alpha \rightarrow 1$ from below. 
Bottom: Decay time $\tau/\tau_0$ as a function of $\alpha$.}
    \label{fig:supp_fig2}
    \end{figure}
%

%
 \subsection{Domain and vortex spectra}
\label{subsec:Spectra}
In \Fig{Three} of the main text we discussed the microscopic origin of characteristic power-laws in the momentum distribution $n(k)$ by decomposing the spectra in different components. 
This was successful since signatures of different defects show up in specific components of the spectra. 
Here we make this correspondence explicit for the case of vortices and domain walls.
 
Under the assumption of spatially constant total particle density the spectrum $n_s$ can be related to the Fourier transform of the correlation function of the angle-averaged spin order parameter $\mathcal{S}(k) = \int \! \mathrm{d}\Omega_k \, \langle S^a(-\vector{k})S^a(\vector{k})\rangle$, 
%
	\begin{align}
	\label{eq:strucfunc}
	\nonumber k^{2}n_s &= \frac{\rho_T}{2} \int \! \mathrm{d}\Omega_k \, \langle \mathcal{F}[\nabla S^{a\ast}] \mathcal{F}[\nabla S^a] \rangle \\
			    &= \frac{\rho_T}{2} k^2 \int \! \mathrm{d}\Omega_k \, \langle S^a(-\vector{k}) S^a(\vector{k}) \rangle,  
	\end{align}     
%
with $\mathcal{F}$ denoting the Fourier transform, and therefore $n_s(k) = \rho_T\mathcal{S}(k)/2$. 
Hence, in a momentum regime ${\pi}/{L_D} \ll k \ll {\pi}/{\xi_s}$ where the the scaling behaviour is dominated by a single domain wall, $n_s \sim \mathcal{S} \sim k^{-3}$ follows from an ansatz in terms of the Heaviside function $S^z = \pm(1 - 2\theta(x)),~S^x \equiv S^y \equiv 0$. 
This feature is similar to the scaling induced by solitons in one-dimensional Bose gases, where the phase jump occurs in the bosonic field and induces a scaling $n_{1D}\sim k^{-2}$, see Ref. 
\cite{Schmidt:2012kw}.

Similarly, the appearance of quantised vortices induces scaling behaviour of the incompressible energy $n_i$. 
On scales considerably smaller than the mean distance between vortices, the effective velocity field $|\mathbf{w}_i(r)|$ associated with the rotational flow decays as $1/r$ with growing distance $r$ from the nearest vortex core. 
As a result, the angle-averaged incompressible kinetic energy distribution $\sim |\mathbf{w}_i^{2}|/2$ gives rise to the IR momentum spectrum $n_i(k)\sim k^{-4}$, whereas compressible excitations vanish.

%
 \subsection{Domain decay near the transition point}
\label{subsec:DomainDecay}
As discussed in the main text, the transition point $\alpha=1$ provides a particularly interesting effect. 
Despite their energetic instability domain walls constitute the relevant spatial pattern associated with the non-thermal fixed point with a scaling $n(k)\sim k^{-3}$. 
For further consideration, we show that this is due to a diverging domain decay time $\tau$ as $\alpha \rightarrow 1$ from below. 
To show this, we plot the time evolution of the integrated density imbalance $\langle |S^z| \rangle= \int \mathrm{d}x^2 \, |S^z|/L^2$ as a function of time in \Fig{supp_fig2} (top left) and $\langle |S^z| \rangle$ at final time $\tau_0$ in \Fig{supp_fig2} (top right). 
From this data, we can extract a decay time $\tau$ for domain structures defined by $\tau=\tau_0/|\langle |S^z| \rangle_f-\langle |S^z| \rangle_i|$ which is shown, on a double-logarithmic scale, in \Fig{supp_fig2} (bottom). 
This definition of the decay time coincides with the first-order term in the Taylor expansion of an exponential decay but has the advantage to be applicable to non-exponential decays. 
One observes a divergence of $\tau$ as $\alpha$ approaches the critical value $\alpha=1$ separating miscible and immiscible regimes. 
This opens interesting possibilities for studying critical exponents in a far-from-equilibrium phase transition~\cite{Karl2012b}.

\end{document}